\documentclass[aps,prl,showpacs,showkeys,
               twocolumn,superscriptaddress,altaffilletter,
               amsmath,amssymb]{revtex4-1}
\usepackage{graphicx}% Include figure files
\usepackage{dcolumn}% Align table columns on decimal point
\usepackage{bm}% bold math
\usepackage{upgreek}
\usepackage{float}
\usepackage{hyperref}% add hypertext capabilities
\usepackage[mathlines]{lineno}% numbering of text and display math
\usepackage{amssymb}     % packages called in main file, e.g. gstr-template.tex
\usepackage{marvosym}    % symbols for EURO
\usepackage{upgreek}     % for upright greek letters in units
\newcommand{\ctsper}      {\ensuremath{\text{cts}/(\text{keV}\cdot\text{kg}\cdot\text{yr})}}

\newcommand{\qbb}         {{$Q_{\beta\beta}$}}

\newcommand{\Qbb}         {\ensuremath{Q_{\beta\beta}}}
\newcommand{\thalfzero}   {\ensuremath{T^{0\nu}_{1/2}}}
\newcommand{\invhl}       {\ensuremath{{\cal S}}}

\newcommand{\onbb}        {\ensuremath{0\nu\beta\beta}}

\newcommand{\gerda}       {\textsc{Gerda}}

\newcommand{\gesix}       {{$^{76}$Ge}}

\newcommand{\bii}         {\ensuremath{\text{BI}_{i}}}
\newcommand{\bisub}          {\ensuremath{\text{BI}_{1}}}

\newcounter{exfig}

\begin{document}
\title{Background free search for neutrinoless double beta decay with
       {\sc Gerda} Phase~II}

\affiliation{INFN Laboratori Nazionali del Gran Sasso and Gran Sasso Science Institute, Assergi, Italy}
\affiliation{INFN Laboratori Nazionali del Sud, Catania, Italy}
\affiliation{Institute of Physics, Jagiellonian University, Cracow, Poland}
\affiliation{Institut f{\"u}r Kern- und Teilchenphysik, Technische Universit{\"a}t Dresden, Dresden, Germany}
\affiliation{Joint Institute for Nuclear Research, Dubna, Russia}
\affiliation{European Commission, JRC-Geel, Geel, Belgium}
\affiliation{Max-Planck-Institut f{\"u}r Kernphysik, Heidelberg, Germany}
\affiliation{Dipartimento di Fisica, Universit{\`a} Milano Bicocca, Milan, Italy}
\affiliation{INFN Milano Bicocca, Milan, Italy}
\affiliation{Dipartimento di Fisica, Universit{\`a} degli Studi di Milano e INFN Milano, Milan, Italy}
\affiliation{Institute for Nuclear Research of the Russian Academy of Sciences, Moscow, Russia}
\affiliation{Institute for Theoretical and Experimental Physics, Moscow, Russia}
\affiliation{National Research Centre ``Kurchatov Institute'', Moscow, Russia}
\affiliation{Max-Planck-Institut f{\"ur} Physik, Munich, Germany}
\affiliation{Physik Department and Excellence Cluster Universe, Technische  Universit{\"a}t M{\"u}nchen, Germany}
\affiliation{Dipartimento di Fisica e Astronomia dell{`}Universit{\`a} di Padova, Padua, Italy}
\affiliation{INFN  Padova, Padua, Italy}
\affiliation{Physikalisches Institut, Eberhard Karls Universit{\"a}t T{\"u}bingen, T{\"u}bingen, Germany}
\affiliation{Physik Institut der Universit{\"a}t Z{\"u}rich, Z{u}rich, Switzerland}
%%%%%%%%%%%%%%%%%%%%%%%%%%%%%%%%%%%%%%%%%%%%%%%%%%%%%%%%%%%%%%%%%%%%%%%%%%%%%%%%

\collaboration{\textsc{Gerda} collaboration}
\email{Correspondence: gerda-eb@mpi-hd.mpg.de}
\noaffiliation
\author{M.~Agostini}
  \affiliation{INFN Laboratori Nazionali del Gran Sasso and Gran Sasso Science Institute, Assergi, Italy}
\author{M.~Allardt}
  \affiliation{Institut f{\"u}r Kern- und Teilchenphysik, Technische Universit{\"a}t Dresden, Dresden, Germany}
\author{A.M.~Bakalyarov}
  \affiliation{National Research Centre ``Kurchatov Institute'', Moscow, Russia}
\author{M.~Balata}
  \affiliation{INFN Laboratori Nazionali del Gran Sasso and Gran Sasso Science Institute, Assergi, Italy}
\author{I.~Barabanov}
  \affiliation{Institute for Nuclear Research of the Russian Academy of Sciences, Moscow, Russia}
\author{L.~Baudis}
  \affiliation{Physik Institut der Universit{\"a}t Z{\"u}rich, Z{u}rich, Switzerland}
\author{C.~Bauer}
  \affiliation{Max-Planck-Institut f{\"u}r Kernphysik, Heidelberg, Germany}
\author{E.~Bellotti}
  \affiliation{Dipartimento di Fisica, Universit{\`a} Milano Bicocca, Milan, Italy}
  \affiliation{INFN Milano Bicocca, Milan, Italy}
\author{S.~Belogurov}
  \affiliation{Institute for Theoretical and Experimental Physics, Moscow, Russia}
  \affiliation{Institute for Nuclear Research of the Russian Academy of Sciences, Moscow, Russia}
\author{S.T.~Belyaev}
 \altaffiliation{deceased}
  \affiliation{National Research Centre ``Kurchatov Institute'', Moscow, Russia}
% ...

%\collaboration{\textsc{Gerda} collaboration}
%\email{Correspondence: gerda-eb@mpi-hd.mpg.de}
%\noaffiliation

\date{\today}

\begin{abstract}
 The Standard Model of particle physics cannot explain the dominance of matter
 over anti-matter in our Universe. In many model extensions this is a very
 natural consequence of neutrinos being their own anti-particles (Majorana
 particles) which implies that a lepton number violating radioactive decay
 named neutrinoless double beta ($0\nu\beta\beta$) decay should exist.  The
 detection of this extremely rare hypothetical process requires utmost
 suppression of any kind of backgrounds.

 The \gerda\ collaboration searches for $0\nu\beta\beta$ decay of $^{76}$Ge
 ($^{76}\rm{Ge} \rightarrow\,^{76}\rm{Se} + 2e^-$) by operating bare detectors
 made from germanium with enriched $^{76}$Ge fraction in liquid argon. Here,
 we report on first data of \gerda\ Phase~II. A background level of
 $\approx10^{-3}$~\ctsper\ has been achieved which is the world-best if
 weighted by the narrow energy-signal region of germanium detectors.
 Combining Phase~I and II data we find no signal and deduce a new lower limit
 for the half-life of $5.3\cdot10^{25}$~yr at 90\,\% C.L.  Our sensitivity of
 $4.0\cdot10^{25}$~yr is competitive with the one of experiments with
 significantly larger isotope mass.

 \gerda\ is the first $0\nu\beta\beta$ experiment that will be background-free
 up to its design exposure.  This progress relies on a novel active veto
 system, the superior germanium detector energy resolution and the improved
 background recognition of our new detectors.  The unique discovery potential
 of an essentially background-free search for $0\nu\beta\beta$ decay motivates
 a larger germanium experiment with higher sensitivity.
\end{abstract}

\vfill

\maketitle
%%%%%%%%%%%%%%%%%%%%%%%%%%%%%%%%%%%%%%%%%%%%%%%%%%%%%%%%%%%%%%%%%%%%%%%%%%%%%%%%
\section{Introduction}

One of the most puzzling aspects of cosmology is the unknown reason for the
dominance of matter over anti-matter in our Universe. Within the Standard
Model of particle physics there is no explanation for this observation and
hence a new mechanism has to be responsible.  A favored model called
leptogenesis~\cite{davidson} links the matter dominance to the nature of
neutrinos and to the violation of lepton number, i.e. the total number of
electrons, muons, taus and neutrinos minus the number of their anti-particles.

In most extensions of the Standard
Model~\cite{mohapatra06,mohapatra07,rodejohann15} neutrinos are assumed to be
their own anti-particles (Majorana particles). This might lead to lepton
number violating processes at the TeV energy scale observable at the
LHC~\cite{rodejohann15} and would result in neutrinoless double beta
($0\nu\beta\beta$) decay where a nucleus of mass number $A$ and charge $Z$
decays as $(A,Z) \rightarrow (A,Z+2) + 2\,e^-$. Lepton number violation has
not been unambiguously observed so far.  There are several experimental
$0\nu\beta\beta$ decay programs ongoing using for example
$^{76}$Ge~\cite{gerda:2013:prl,2015:mjd},
$^{130}$Te~\cite{2015:cuore,2016:sno} or
$^{136}$Xe~\cite{kamland:2016,exo:2014,next:2016}.  They all measure the sum
of the electron energies released in the decay which corresponds to the mass
difference $Q_{\beta\beta}$ of the two nuclei. The $0\nu\beta\beta$ decay
half-life is at least 15 orders of magnitude longer than the age of the
universe. Its observation requires therefore the best suppression of
backgrounds.

In the GERmanium Detector Array (\gerda) experiment bare germanium detectors
are operated in liquid argon (LAr).  The detectors are made from germanium
with the $^{76}$Ge isotope fraction enriched from 7.8\,\% to about 87\,\%.
Since source and detector of $0\nu\beta\beta$ decay are identical in this
calorimetric approach the detection efficiency is high.

This Article presents the first result from \gerda\ Phase~II.  In the first
phase of data taking (Phase~I), a limit of $T_{1/2}^{0\nu}>2.1\cdot10^{25}$~yr
(90\,\% C.L.)  was found~\cite{gerda:2013:prl} for an exposure of
21.6~kg$\cdot$yr and a background of 0.01~\ctsper\ at
$Q_{\beta\beta}=(2039.061\pm0.007)$~keV~\cite{qbb}.  At that time, the result
was based on data from 10 detectors (17.6~kg total mass).  In December 2015,
Phase~II started with 37 detectors (35.6~kg) from enriched material.  The mass
is hence doubled relative to Phase~I. The ambitious goal is an improvement of
the half-life sensitivity to $>10^{26}$~yr for about 100 kg$\cdot$yr exposure
by reducing the background level by an order of magnitude. The latter is
achieved by vetoing background events through the detection of their energy
deposition in LAr and the characteristic time profile of their signals in the
germanium detectors.  The expected background is less than one count in the
energy region of interest up to the design exposure which means that
\gerda\ will be the first ``background free'' experiment in the field.

We will demonstrate in this Article that \gerda\ has reached the envisioned
background level which is the world-best level if weighted by our superior
energy resolution. \gerda\ is therefore best suited to not only quote limits
but to identify with high confidence a $0\nu\beta\beta$ signal.

\section{The experiment}

The \gerda\ experiment~\cite{gerda:2013:tec} is located at the underground
Laboratori Nazionali del Gran Sasso (LNGS) of INFN, Italy.  A rock overburden
of about 3500~m water equivalent removes the hadronic components of cosmic ray
showers and reduces the muon flux at the experiment by six orders of magnitude
to 1.2~$\mu$/(m$^2\cdot$h).

The basic idea is to operate bare germanium detectors in a radiopure cryogenic
liquid like LAr for cooling to their operating temperature of $\sim$90~K and
for shielding against external radiation originating from the walls (see
Extended Data Fig.~\ref{extfig:setup} for a sketch of the
setup)~\cite{heusser}.  In \gerda, a 
64~m$^3$ LAr cryostat is inside a 590~m$^3$ water tank. The clean water
completes the passive shield.  Above the water tank is a clean room with a
glove box and lock for the assembly of germanium detectors into strings and
the integration of the liquid argon veto system.

\gerda\ deploys 7 coaxial detectors from the former
Heidelberg-Moscow~\cite{klapdor1} and IGEX~\cite{igex} experiments and 30
broad energy (BEGe) detectors ~\cite{gerda:2015:bege}.  All diodes have p-type
doping (see Extended Data Fig.~\ref{extfig:detectors}). Electron-hole pairs
created in the 1--2~mm thick n$+$ electrode mostly recombine such that the
active volume is reduced.  A superior identification of the event topology and
hence background rejection is available for the BEGe type (see below).  The
enriched detectors are assembled into 6 strings surrounding the central one
which consists of three coaxial detectors of natural isotopic composition.
Each string is inside a nylon cylinder (see Extended Data
Fig.~\ref{extfig:string}) to limit the LAr volume from which radioactive ions
like $^{42}$K can be collected to the outer detector
surfaces~\cite{gerda:2014:bkg}.

All detectors are connected to custom made low radioactivity charge sensitive
amplifiers~\cite{cc3} (30~MHz bandwidth, 0.8~keV full width at half maximum
(FWHM) resolution) located in LAr about 35~cm above the detectors.  The charge
signal traces are digitized with 100~MHz sampling rate and stored on disk for
offline analysis.

In background events some energy is often also deposited in the argon. The
resulting scintillation light~\cite{lar1} can be detected to veto them.  In
Phase~II, a cylindrical volume of 0.5~m diameter and 2.2~m height around the
detector strings (see Extended Data Fig.~\ref{extfig:setup}
and~\ref{extfig:larcaps}) is instrumented with light sensors.  The central
0.9~m of the cylinder are defined by a curtain of wavelength shifting fibers
which surround the 0.4~m high detector array. The fibers are read-out at both
ends with 90~silicon photomulipliers (SiPM)~\cite{lar2}.  Groups of six
$3\times3$~mm$^2$ SiPMs are connected together to a charge sensitive
amplifier.  Sixteen 3'' low-background photomultpliers (PMT) designed for
cryogenic operation are mounted at the top and bottom surfaces of the
cylindrical volume. The distance to any detector is at least 0.7~m to limit
the PMT background contribution from their intrinsic Th/U radioactivity.  All
LAr veto channels are digitized and read-out together with the germanium
channels if at least one detector has an energy deposition above
$\sim$100~keV.

The nylon cylinders, the fibers, the PMTs and all surfaces of the instrumented
LAr cylindrical volume are covered with a wavelength shifter to shift the LAr
scintillation light from 128~nm to about 400~nm to match the peak quantum
efficiency of the PMTs and the absorption maximum of the fibers.

The water tank is instrumented with 66~PMTs to detect Cherenkov light from
muons passing through the experiment.  On top of the clean room are three
layers of plastic scintillator panels covering the central 4$\times$3~m$^2$ to
complete the muon veto~\cite{gerda:muon}.

\section{Data analysis}

The data analysis flow is very similar to that of Phase~I.  The offline
analysis of the digitized germanium signals is described in
Refs.~\cite{gerda:2013:prl,acat,gelatio}.

A data blinding procedure is again applied.  Events with a reconstructed
energy in the interval $Q_{\beta\beta}\pm 25$~keV are not analyzed but only
stored on disk.  After the entire analysis chain has been frozen, these
blinded events have been processed.

The gain stability of each germanium detector is continuously monitored by
injecting charge pulses (test pulses) into the front-end electronics with a
rate of 0.05~Hz. The test pulses are also used to monitor leakage current and
noise. Only data recorded during stable operating conditions (e.g.~gain
stability better than 0.1\,\%) are used for the physics analysis. This
corresponds to about 85\,\% of the total data written on disk.

Signals originated from electrical discharges in the high voltage line or
bursts of noise are rejected during the off\-line event reconstruction by a
set of multi-parametric cuts based on the flatness of the baseline, polarity
and time structure of the pulse. Physical events at $Q_{\beta\beta}$ are
accepted with an efficiency larger than 99.9\,\% estimated with $\gamma$ lines
in calibration data, test pulse events and template signals injected in the
data set.  Conversely, a visual inspection of all events above 1.6~MeV shows
that no unphysical event survives the cuts.

\begin{figure}
\includegraphics[width=\columnwidth]{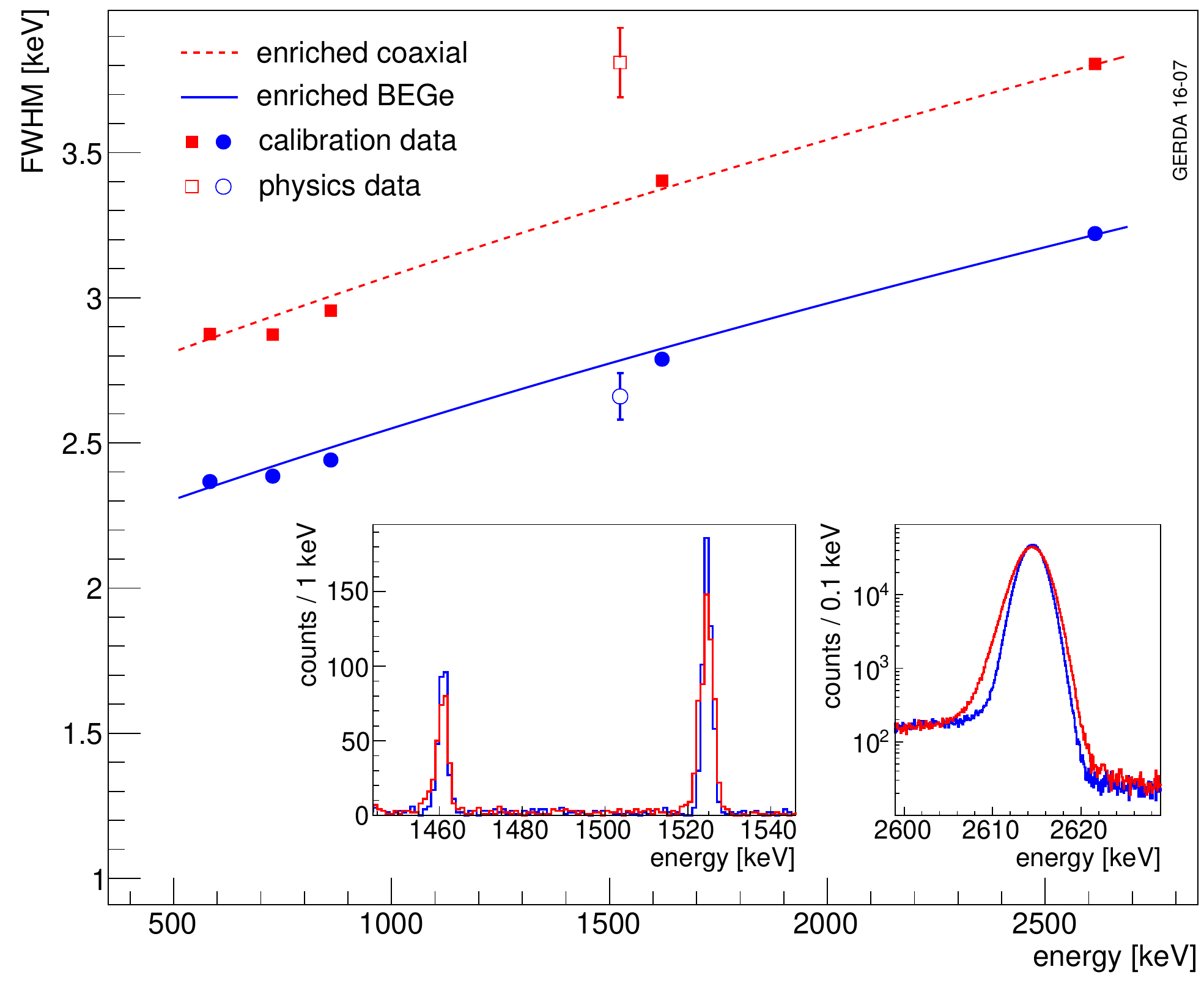}
\caption{\label{fig:Eres}
        Average energy resolution (FWHM) for $\gamma$ lines of the calibration
        spectrum (filled symbols) and the $^{42}$K line from physics data
        (open symbols) for BEGe (symbols and solid line in blue) and coaxial
        (symbols and dashed line in red) detectors. The insets show the K
        lines and the 2615~keV calibration peak.
}
\end{figure}

The energy deposited in a germanium detector is reconstructed offline with an
improved digital filter~\cite{gerda:2015:zac}, whose parameters are optimized
for each detector and for several periods.  The energy scale and resolution
are determined with weekly calibration runs with $^{228}$Th sources.  The
long-term stability of the scale is assessed by monitoring the shift of the
position of the 2615~keV peak between consecutive calibrations. It is
typically smaller than 1~keV for BEGe detectors and somewhat worse for some
coaxial ones.  The FWHM resolution at 2.6~MeV is between 2.6--4.0~keV for BEGe
and 3.4--4.4~keV for coaxial detectors. The width of the strongest $\gamma$
lines in the physics data (1525~keV from $^{42}$K and 1460~keV from $^{40}$K)
is found to be 0.5~keV larger than the expectation for the coaxial detectors
(see Fig.~\ref{fig:Eres}).  In order to estimate the expected energy
resolution at $Q_{\beta\beta}$ an additional noise term is added to take this
into account.

For $0\nu\beta\beta$ decays in the active part of a detector volume, the total
energy of $Q_{\beta\beta}$ is detected in 92\,\% of the cases in this
detector.  Multiple detector coincidences are therefore discarded as
background events.  Two consecutive candidate events within 1~ms are also
rejected (dead time $\sim$$10^{-4}$) to discriminate time-correlated decays
from primordial radioisotopes, as e.g.~the radon progenies $^{214}$Bi and
$^{214}$Po.  Candidate events are also refuted if a muon trigger occurred
within 10~$\upmu$s prior to a germanium detector trigger. More than 99\,\% of
the muons that deposit energy in a germanium detector are rejected this way.
The induced dead time is $<$0.1\,\%.

The traces from PMTs and SiPMs are analyzed offline to search for LAr
scintillation signals in coincidences with a germanium detector trigger. An
event is rejected if any of the light detectors record a signal of amplitude
above 50\,\% of the expectation for a single photo-electron within 5~$\upmu$s
from the germanium trigger.  99\,\% of the photons occur in this window.
Accidental coincidences between the LAr veto system and germanium detectors
create a dead time of $(2.3\pm0.1)$\,\% which is measured with test pulse
events and cross checked with the counts in the $^{40}$K peak.

\begin{figure*}
\includegraphics[width=\textwidth]{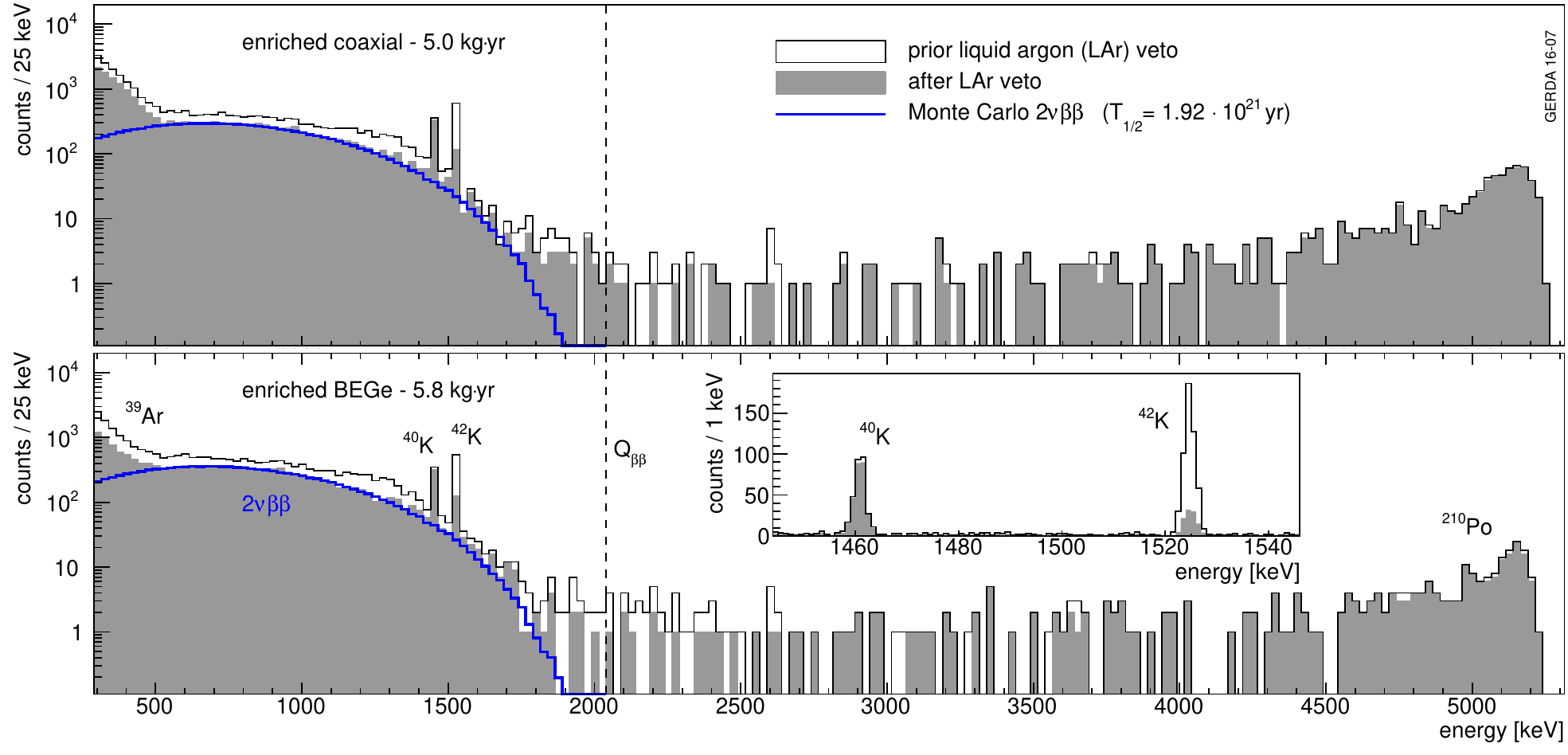}
\caption{\label{fig:allE}
         Energy spectra of Phase~II data sets before (open histogram) and
         after argon veto cut (filled histogram). The blue lines are the
         expected $2\nu\beta\beta$ spectra for our recent half-life
         measurement. The inset shows the BEGe spectrum in the energy region
         around the two potassium lines. Various background contributions are
         labeled in the bottom panel.
}
\end{figure*}

Fig.~\ref{fig:allE} shows the energy spectra for BEGe and coaxial detectors of
Phase~II with and without the LAr veto cut.  Below $\sim$$500$~keV the spectra
are dominated by $^{39}$Ar $\beta$ decays, up to 1.7~MeV by events from double
beta decay with two neutrino emission ($2\nu\beta\beta$), above 2.6~MeV by
$\alpha$ decays on the detector surface and around $Q_{\beta\beta}$ by a
mixture of $\alpha$ events, $^{42}$K $\beta$ decays and those from the
$^{238}$U and $^{232}$Th decay chains. The two spectra are similar except for
the number of $\alpha$ events which is on average higher for coaxial
detectors.  The number of $\alpha$ counts shows a large variation between the
detectors.  The power of the LAr veto is best demonstrated by the $^{42}$K
line at 1525~keV which is suppressed by a factor $\sim$5 (see inset) due to
the $\beta$ particle depositing up to 2~MeV energy in the LAr.  The figure
also shows the predicted $2\nu\beta\beta$ spectrum from $^{76}$Ge using our
Phase~I result for the half-life of
$T_{1/2}^{2\nu}=(1.926\pm0.094)\cdot10^{21}$~yr~\cite{gerda:2015:2nbb}.

The time profile of the germanium detector current signal is used to
discriminate $0\nu\beta\beta$ decays from background events.  While the former
have point-like energy deposition in the germanium (single site events, SSE),
the latter have often multiple depositions (multi site events, MSE) or
depositions on the detector surface. The same pulse shape discrimination (PSD)
techniques of Phase~I~\cite{gerda:2013:psd} are applied.

Events in the double escape peak (DEP) and at the Compton edge of 2615~keV
gammas in calibration data have a similar time profile as $0\nu\beta\beta$
decays and are hence proxies for SSE. These samples are used to define the PSD
cuts and the related detection efficiencies.  The latter are cross checked
with $2\nu\beta\beta$ decays.

\begin{figure}
\includegraphics[width=\columnwidth]{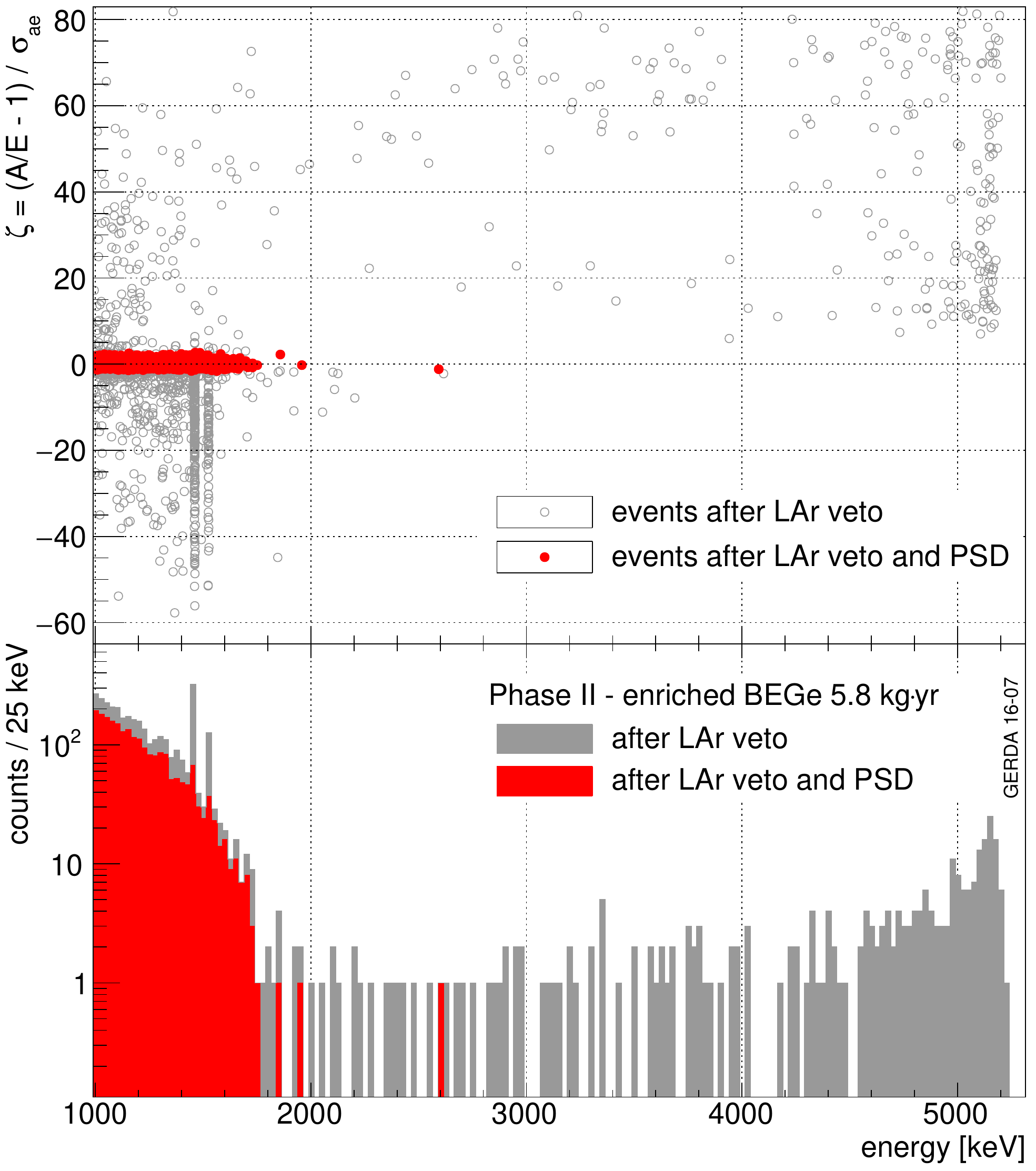}
\caption{\label{fig:aovere} 
                  For all BEGe detectors, PSD parameter $\zeta = (A/E -
                  1)/\sigma_{ae}$ versus energy for physics data and the
                  projection to the energy axis. Red circles and red spectrum
                  represent events that pass the selection.  Since the cuts
                  are detector specific the accepted $\zeta$ ranges differ.
}
\end{figure}

The geometry of BEGe detectors allows to apply a simple mono-parametric PSD
based on the maximum of the detector current pulse $A$ normalized to the total
energy $E$~\cite{aovere,aovere2}.  The energy dependence of the mean and the
resolution $\sigma_{ae}$ of $A/E$ are measured for every detector with
calibration events. After correcting for these dependences and normalizing the
mean $A/E$ of DEP events to~1, the acceptance range is determined for each
detector individually: the lower cut is set to keep 90\,\% of DEP events and
the upper position is twice the low-side separation
from~1. Fig.~\ref{fig:aovere} shows a scatter plot of the PSD parameter
$\zeta=(A/E-1)/\sigma_{ae}$ versus energy and the projection to the energy
axis. Events marked in red survive the PSD selection.  Below 1.7~MeV
$2\nu\beta\beta$ events dominate with a survival fraction of
$(85_{-1}^{+2})$\,\%.  The two potassium peaks and Compton scattered photons
reconstruct at $A/E<1$ (below the SSE band).  All 234~$\alpha$ events at
higher energies exhibit $A/E > 1$ and are easily removed.  The average
$0\nu\beta\beta$ survival fraction~\cite{phd:vici} is $(87\pm2)$\,\%. The
uncertainty takes into account the systematic difference between the $A/E$
centroids of DEP and $2\nu\beta\beta$ events and different fractions of MSE in
DEP and $0\nu\beta\beta$ events.

For coaxial detectors a mono-parametric PSD is not sufficient since SSE do not
have a simple signature~\cite{gerda:2013:psd}.  Instead two neural network
algorithms are applied to discriminate SSE from MSE and from $\alpha$ surface
events. The first one is identical to the one used in Phase~I.  The cut on the
neural network qualifier is set to yield a survival fraction of DEP events of
90\,\% for each detector. For the determination of the $0\nu\beta\beta$
efficiency, $2\nu\beta\beta$ events in physics data and a complete Monte Carlo
simulation~\cite{kirschphd} of physics data and calibration data are used. The
simulation considers the detector and the electronics response to energy
depositions including the drift of charges in the crystal~\cite{adl}.  We find
a survival fraction for $0\nu\beta\beta$ events of $(85\pm5)$\,\% where the
error is derived from variations of the simulation parameters.

The second neural network algorithm is applied for the first time and
identifies surface events on the p$+$ contact.  Training is done with physics
data from two different energy intervals. After the LAr veto cut events in the
range 1.0--1.3~MeV are almost exclusively from $2\nu\beta\beta$ decay and hence
signal-like.  Events above 3.5~MeV are almost all from $\alpha$ decays on the
p$+$ electrode and represent background events in the training.  As
$0\nu\beta\beta$ efficiency we measure a value of $(93\pm1)$\,\% for a
$2\nu\beta\beta$ event sample not used in the training.  The combined PSD
efficiency for coaxial detectors is $(79\pm5)$\,\%.

\section{Results}

This analysis includes the data sets used in the previous
publication~\cite{gerda:2013:prl,gerda:2015:taup13}, an additional coaxial
detector period from 2013 (labeled ``PI extra'') and the Phase~II data from
December 2015 until June 2016 (labeled ``PIIa coaxial'' and ``PIIa
BEGe''). Table~\ref{tab:datasets} lists the relevant parameters for all data
sets.  The exposures in the active volumes of the detectors for $^{76}$Ge are
234 and 109 mol$\cdot$yr for Phase~I and II, respectively.  The
efficiency is the product of the $^{76}$Ge isotope fraction (87\,\%), the
active volume fraction (87--90\,\%), the $0\nu\beta\beta$ event fraction
reconstructed at full energy in a single crystal (92\,\%), pulse shape
selection (79--92\,\%) and the live time fraction (97.7\,\%).  For the Phase~I
data sets the event selection including the PSD classification is
unchanged. An improved energy reconstruction~\cite{gerda:2015:zac} is applied
to the data as well as an updated value for the coaxial detector PSD
efficiency of the neural network analysis of $(83\pm3)$\,\%~\cite{kirschphd}.

\begin{table}
\caption{\label{tab:datasets}
        List of data sets, exposures (for total mass), energy resolutions in
        FWHM, efficiencies (including enrichment, active mass, reconstruction
        efficiencies and dead times) and background indices (BI) in the analysis
        window.
}
\begin{tabular}{ccccc} \hline
data set  & exposure & FWHM & efficiency &  BI \\ 
          &  [kg$\cdot$yr] & [keV] &  & $10^{-3}$\ctsper \\ \hline
PI golden & 17.9 & $4.3(1)$ & $0.57(3)$ & $11\pm2$~~  \\
PI silver & 1.3  & $4.3(1)$ & $0.57(3)$ & $30\pm10$  \\
PI BEGe   & 2.4  & $2.7(2)$ & $0.66(2)$ & $5_{-3}^{+4}$ \\
PI extra  & 1.9  & $4.2(2)$ & $0.58(4)$ & $5_{-3}^{+4}$ \\ \hline
PIIa coaxial & 5.0  & $4.0(2)$ & $0.53(5)$ & $3.5_{-1.5}^{+2.1}$ \\
PIIa BEGe & 5.8  & $3.0(2)$ & $0.60(2)$ & $0.7_{-0.5}^{+1.1}$ \\ \hline
\end{tabular}
\end{table}

\begin{figure}
\includegraphics[width=\columnwidth]{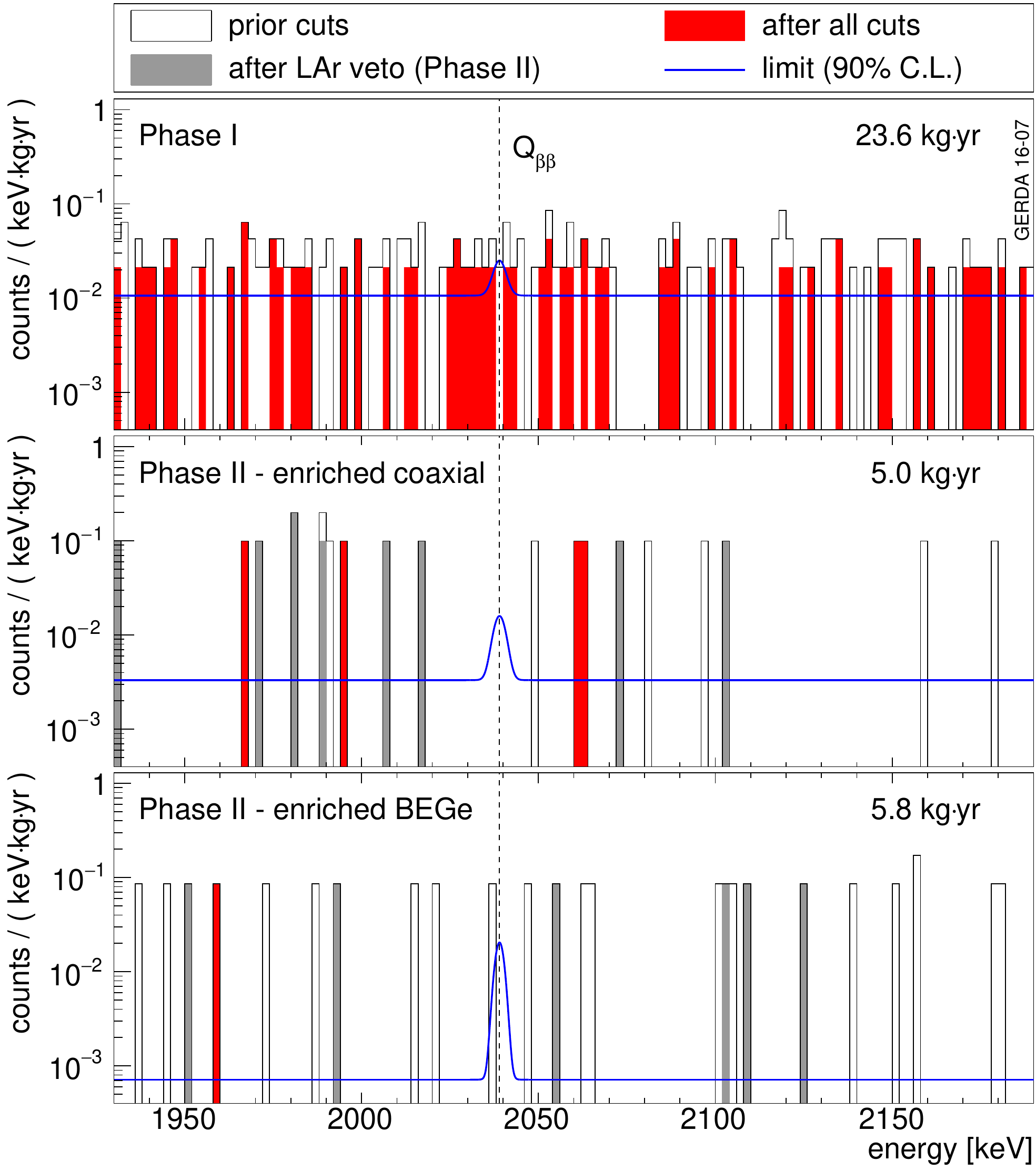}
\caption{\label{fig:spectrum}
         Combined Phase~I data (top), Phase~II coaxial (middle) and BEGe
         detector spectra (bottom) in the analysis window. The binning is
         2~keV. The exposures are given in the panels. The red histogram is
         the final spectrum, the filled grey one without pulse shape
         discrimination and the open one in addition without argon veto cut.
         The blue line is the fitted spectrum together with a hypothetical
         signal corresponding to the 90\,\% C.L. limit of $T_{1/2}^{0\nu} >
         5.3\cdot10^{25}$~yr. 
}
\end{figure}

Fig.~\ref{fig:spectrum} shows the spectra for the combined Phase~I data sets
and the two Phase~II sets.  The analysis range is from 1930 to 2190~keV
without the intervals $(2104\pm5)$~keV and $(2119\pm5)$~keV of known peaks
predicted by our background model~\cite{gerda:2014:bkg}.  For the coaxial
detectors four events survive the cuts which means that the background is
reduced by a factor of three compared to Phase~I (see 'PI golden' in
Tab.~\ref{tab:datasets}).  Due to the better PSD performance, only one event
remains in the BEGe data which corresponds to a background of
$0.7_{-0.5}^{+1.1}\cdot 10^{-3}$ \ctsper.  Consequently, the Phase~II
background goal is reached.

We perform both a Frequentist and a Bayesian analysis based on an unbinned
extended likelihood function~\cite{gerda:2015:taup13}. The fit function for
every data set is a flat distribution for the background (one free parameter
per set) and for a possible signal a Gaussian centered at $Q_{\beta\beta}$
with a width according to the corresponding resolution listed in
Tab.~\ref{tab:datasets}. The signal strength is calculated for each set
according to its exposure, efficiency and the inverse half-life $1/T$ which is
a common free parameter.

Systematic uncertainties like a 0.2~keV uncertainty of the energy scale at
$Q_{\beta\beta}$ are included in the analysis as pull terms in the likelihood
function.  The implementation takes correlations into account.

The Frequentist analysis uses the Neyman construction of the confidence
interval and the standard two-sided test statistics~\cite{pdg14,cowan} with
the restriction to the physical region $1/T\ge0$: the frequency distribution
of the test statistic is generated using Monte Carlo simulations for different
assumed $1/T$ values.  The limit was determined by finding the largest value
of $1/T$ for which at most 10\,\% of the simulated experiments had a value of
the test statistic more unlikely than the one measured in our data (see
Extended Data Fig.~\ref{extfig:analysis}).  Details of the statistical
analysis can be found in the appendix. The best fit yields zero signal events
and a 90\,\% C.L.~limit of 2.0 events in 34.4~kg$\cdot$yr total exposure or
\begin{equation}
T_{1/2}^{0\nu} > 5.3 \cdot 10^{25}\,{\rm yr.}
\end{equation}
The (median) sensitivity assuming no signal is $4.0\cdot10^{25}$~yr (see
Extended Data Fig.~\ref{extfig:analysis}). The systematic errors weaken the
limit by $<$1\,\%.

The Bayesian fit yields for a prior flat in $1/T$ between 0 and
$10^{-24}$~yr$^{-1}$ a limit of $T_{1/2}^{0\nu} > 3.5\cdot10^{25}$~yr (90\,\%
C.I.).  The sensitivity assuming no signal is $3.1\cdot10^{25}$~yr.

\section{Discussion}

The second phase of \gerda\ collects data since December 2015 in stable
conditions with all channels working.  The background at $Q_{\beta\beta}$ for
the BEGe detectors is $(0.7_{-0.5}^{+1.1})\cdot10^{-3}$~\ctsper. This is a
major achievement since the value is consistent with our ambitious design
goal.

We find no hint for a $0\nu\beta\beta$ decay signal in our combined data and
place a limit of $T_{1/2}^{0\nu} ({\rm ^{76}Ge})>5.3\cdot10^{25}$~yr (90\,\%
C.L., sensitivity $4.0\cdot10^{25}$~yr).  For light Majorana neutrino exchange
and a nuclear matrix element range for $^{76}$Ge between 2.8 and 6.1
\cite{mene09,horoi16,bar15,suh15,fae13,rod13,yao15} the \gerda\ half-life
limit converts to $m_{\beta\beta}<$0.15--0.33~eV (90\,\% C.L.).

We expect only a fraction of a background event in the energy region of
interest (1 FWHM) at design exposure of 100~kg$\cdot$yr. \gerda\ is hence the
first ``background free'' experiment in the field. Our sensitivity grows
therefore almost linearly with time instead of by square root like for
competing experiments and reaches $10^{26}$~yr for the half-life limit within
3 years of continuous operation. With the same exposure we have a 50\,\% chance
to detect a signal with $3\sigma$ significance if the half-life is below
$10^{26}$~yr.

Phase~II has demonstrated that the concept of background suppression by
exploiting the good pulse shape performance of BEGe detectors and by detecting
the argon scintillation light works.  The background at $Q_{\beta\beta}$ is at
a world-best level: it is lower by typically a factor of 10 compared to
experiments using other isotopes after normalization by the energy resolution
and total efficiency $\epsilon$; i.e. (BI$\cdot$FWHM/$\epsilon$) is superior.
This is the reason why the \gerda\ half-life sensitivity of
$4.0\cdot10^{25}$~yr for an exposure of 343~mol$\cdot$yr is similar to the one
of Kamland-Zen for $^{136}$Xe of $5.6\cdot10^{25}$~yr based on a more than
10-fold exposure of 3700~mol$\cdot$yr~\cite{kamland:2016}.

A discovery of $0\nu\beta\beta$ decay would have far reaching consequences for
our understanding of particle physics and cosmology. Key features for a
convincing case are an ultra low background with a simple flat distribution,
excellent energy resolution and the possibility to identify the events with
high confidence as signal-like as opposed to an unknown $\gamma$-line from a
nuclear transition.  The latter is achieved by the detector pulse shape
analysis and possibly a signature in the argon.  The concept to operate bare
germanium detectors in liquid argon has proven to have the best performance
for a discovery which motivates future extensions of the program. The
\gerda\ cryostat can hold 200~kg of detectors. Such an experiment will remain
background-free until an exposure of 1000~kg$\cdot$yr provided the background
can be further reduced by a factor of five.  The discovery sensitivity would
then improve by an order of magnitude to a half-life of $10^{27}$~yr.  The
200~kg setup is conceived as a first step for a more ambitious 1~ton
experiment which would ultimately boost the sensitivity to $10^{28}$~yr
corresponding to the $m_{\beta\beta}<$10--20~meV range. Both extensions
are being pursued by the newly formed LEGeND Collaboration
(http://www.legend-exp.org)

%%%%%%%%%%%%%%%%%%%%%%%%%%%%%%%%%%%%%%%%%%%%%%%%%%%%%%%%%%%%%%%%%%%%%%%%%%%%%%%%
\appendix
\section{Acknowledgments}       %%%%%%%%%%%%%%%%%%%%%%%%%%%%%%%%%%%%%%%%%%%%%%%%
%\begin{acknowledgments}
 The \gerda\ experiment is supported financially by
   the German Federal Ministry for Education and Research (BMBF),
   the German Research Foundation (DFG) via the Excellence Cluster Universe,
   the Italian Istituto Nazionale di Fisica Nucleare (INFN),
   the Max Planck Society (MPG),
   the Polish National Science Centre (NCN),
   the Russian Foundation for Basic Research (RFBR), and
   the Swiss National Science Foundation (SNF).
 The institutions acknowledge also internal financial support.

The \gerda\ collaboration thanks the directors and the staff of the LNGS
for their continuous strong support of the \gerda\ experiment.

%\end{acknowledgments}
%%%%%%%%%%%%%%%%%%%%%%%%%%%%%%%%%%%%%%%%%%%%%%%%%%%%%%%%%%%%%%%%%%%%%%%%%%%%%%%%
\appendix
\section{Appendix: Statistical Methods }       %%%%%%%%%%%%%%%%%%%%%%%%%%%%%%%%%

This section discusses the statistical analysis of the \gerda\ data. In
particular, the procedures to derive the limit on \thalfzero, the median
sensitivity of the experiment and the treatment of systematic uncertainties
are described.

A combined analysis of data from Phase\,I and~II is performed by fitting
simultaneously the six data sets %$\boldsymbol{\mathcal{D}}$
of Table~\ref{tab:datasets}.  The parameter of interest for this analysis is
the strength of a possible \onbb\ decay signal: $\invhl = 1/\thalfzero$.  The
number of expected \onbb\ events in the $i$-th data set {$\mathcal{D}_i$} as a
function of \invhl\ is given by:
\begin{equation}
 \label{eq:conversion}
 \mu^{S}_i  = \ln 2 \cdot \left(N_{A}/m_{a}\right) \cdot 
     \epsilon_i \cdot {\cal E}_i
 \cdot \invhl\   ,
\end{equation}
where $N_A$ is Avogadro's number, $\epsilon_i$ the global signal efficiency of
the $i$-th data set, ${\cal E}_i$ the exposure and $m_a$ the molar mass.  The
exposure quoted is the total detector mass multiplied by the data taking time.
The global signal efficiency accounts for the fraction of \gesix\ in the
detector material, the fraction of the detector active volume, the efficiency
of the analysis cuts, the fractional live time of the experiment and the
probability that \onbb\ decay events in the active detector volume have a
reconstructed energy at \qbb.  The total number of expected background events
as a function of the background index $\bii$ is:
\begin{equation}
 \label{eq:conversionbackground}
 \mu^{B}_i  = {\cal E}_i \cdot \bii \cdot \Delta E,
\end{equation}
where $\Delta E$=240\,keV is the width of the energy region around \qbb\ used
for the fit.

Each data set $\mathcal{D}_i$ is fitted with an unbinned likelihood function
assuming a Gaussian distribution for the signal and a flat distribution for
the background:
\begin{equation} \label{partiallikelihood}
\begin{split}
   \mathcal{L}_i&(\mathcal{D}_i|\invhl, \bii, \theta_i) \ = \
    \prod_{j=1}^{N^{obs}_i}  \  \dfrac{1}{ \mu^{S}_i +  \mu^{B}_i}\,\cdot\\ 
   & \left[  
      \mu^{S}_i\cdot\dfrac{1}{\sqrt{2 \pi} \sigma_i}% \cdot
      \exp\left(\dfrac{-(E_j-\Qbb - \delta_i)^2}{2 \sigma_i^2}\right) +
      \mu^{B}_i\cdot\dfrac{1}{\Delta E}
                               \right]
\end{split}
\end{equation}
where $E_j$ are the individual event energies, ${N^{obs}_i}$ is the total
number of events observed in the $i$-th data set, $\sigma_i=\text{FWHM}_i/(2
\sqrt{2\ln{2}})$ is the energy resolution and $\delta_i$ is a possible
systematic energy offset.  The parameters with systematic uncertainties are
indicated with $\theta_i = \{\epsilon_i, \sigma_i, \delta_i\}$.  The
parameters $\invhl$ and $\bii$ are bound to positive values.  The total
likelihood is constructed as the product of all $\mathcal{L}_i$ weighted with
the Poisson terms~\cite{pdg}:
\begin{equation} \label{fulllikelihood}
\begin{split}
   \mathcal{L}&(\boldsymbol{\mathcal{D}}|\invhl, \text{\bf{BI}},
   \boldsymbol{\theta}) 
   =\\& \prod_i \left[
      \dfrac{e^{-(\mu^{S}_i + \mu^{B}_i)} \cdot ( \mu^{S}_i + \mu^{B}_i )^{N^{obs}_i}}
      {N^{obs}_i!}
   \cdot \mathcal{L}_i(\mathcal{D}_i|\invhl, \bii, \theta_i)
  \right]
\end{split}
\end{equation}
where $\boldsymbol{\mathcal{D}}=\{\mathcal{D}_1\dots\mathcal{D}_i\dots\}$,
$\text{\bf BI}=\{\bisub\dots\bii\dots\}$ and
$\boldsymbol{\theta}=\{\theta_1\dots\theta_i\dots\}$.

A frequentist analysis is performed using a two-sided test
statistics~\cite{cowan} based on the profile likelihood $\lambda$ (\invhl):
\begin{equation}
\label{eq:t}
t_\invhl = 
-2 \ln {\lambda}(\invhl) =
-2 \ln \frac{\mathcal{L}(\invhl, \hat{\hat{\text{\bf BI}}},
                                     \hat{\hat{\boldsymbol{\theta}}})}
            {\mathcal{L}(\hat{\invhl}, \hat{\text{\bf BI}},
                                         \hat{\boldsymbol{\theta}})}
\end{equation}
where $\hat{\hat{\text{\bf BI}}}$ and $\hat{\hat{\boldsymbol{\theta}}}$ in the
numerator denote the value of the parameters that maximizes $\mathcal{L}$ for
a fixed \invhl.  In the denominator, $\hat{\invhl}$,
$\hat{\boldsymbol{\text{BI}}}$ and $\hat{\boldsymbol{\theta}}$ are the values
corresponding to the absolute maximum likelihood.

The confidence intervals are constructed for a discrete set of values
$\invhl\in\{\invhl_j\}$.  For each $\invhl_j$, possible realizations of the
experiments are generated via Monte Carlo according to the parameters of
Table~\ref{tab:datasets} and the expected number of counts from
Eqs.~\ref{eq:conversion} and~\ref{eq:conversionbackground}.
For each realization  ${t_\invhl}_j$ is evaluated. From the
entire set the probability distribution $f(t_{\invhl}|\invhl_j)$ is calculated.
The p-value of the data for a specific $\invhl_j$
is computed as: 
\begin{equation}  \label{eq:pj_integral}
   p_{\invhl_j} = \int_{t_{obs}}^\infty  f(t_{\invhl}|\invhl_j) \, d({t_{\invhl}}_j)
\end{equation}
where $t_{obs}$ is the value of the test statistics of the \gerda\ data for
$\invhl_j$. The values of $p_{\invhl_j}$ are shown by the solid line in
Extended Data Fig.~\ref{extfig:analysis}.  The 90\,\% C.L.~interval is given
by all $\invhl_j$ values with $p_{\invhl_j}> 0.1$.  Such an interval has the
correct coverage by construction.  The current analysis yields a one-sided
interval, i.e. a limit of $\thalfzero =1/\invhl> 5.3\cdot10^{25}$~yr.

The expectation for the frequentist limit (i.e. the experimental sensitivity)
was evaluated from the distribution of $p_{\invhl_j}$ built from Monte Carlo
generated data sets with no injected signal ($\invhl=0$).  The distribution of
$p_{\invhl_j}$ is shown in Extended Data Fig.~\ref{extfig:analysis}: the
dashed line is the median of the distribution and the color bands indicate the
68\,\% and 90\,\% probability central intervals.  The experimental sensitivity
corresponds to the \invhl\ value at which the median crosses the p-value
threshold of 0.1\,: $\thalfzero>4.0\cdot 10^{25}$~yr (90\% C.L.).

Systematics uncertainties are folded into the likelihood by varying the
parameters $\theta_i$ in the fits and constraining them by adding to the
likelihood multiplicative Gaussian penalty terms.  The central values and the
standard deviations of the penalty terms for $\epsilon_i$ and $\sigma_i$ are
taken from Table~\ref{tab:datasets}.  The penalty term on $\delta_i$ has a
central value equal to zero and standard deviation of 0.2\,keV.

Instead of the two-sided test statistics one can use a one-sided test
statistic defined as~\cite{cowan}:
\begin{equation}
   \tilde{t}_{\invhl} = 
%-2 \ln {\tilde{\lambda}}(\invhl) =
 \: \left\{ \! \! 
    \begin{array}{ll}
       0, 
       & \quad \hat{\invhl} > \invhl\ge0  \\*[0.2 cm]
       -2 \ln {\lambda}(\invhl), 
        & \quad \hat{\invhl} \leq \invhl  \\*[0.2 cm]
     \end{array}
       \right.
\end{equation}
By construction $\tilde{t}_{\invhl}=0$ for $\invhl=0$ for all realizations and
consequently $\invhl= 0$ is always included in the 90\,\% C.L.~interval,
i.e.~the one-sided test statistic will always yield a limit. In our case the
resulting limit would be 50\% stronger.  Similar to other
experiments~\cite{exo:2014,kamland:2016}, we want to be able to detect a
possible signal and thus we decided a priori to adopt the two-sided test
statistic. It is noteworthy that, although the coverage of both test
statistics is correct by construction, deciding which one to use according to
the outcome of the experiment would result in the flip-flop issue discussed by
Feldman and Cousins~\cite{Feldman:1997qc}.

The statistical analysis is also performed within a Bayesian framework. The
combined posterior probability density function (PDF) is calculated from the
six data sets according to Bayes' theorem:
\begin{equation} \label{posterior}
  \mathcal{P}(\invhl,\text{\bf BI}|\boldsymbol{\mathcal{D}},\boldsymbol{\theta})
   \propto \mathcal{L}(\boldsymbol{\mathcal{D}}|\invhl,
   \text{\bf BI}, \boldsymbol{\theta})  \ \mathcal{P}(\invhl)  \ 
   \prod_i \mathcal{P}(\bii) 
\end{equation}
The likelihood $\mathcal{L}$ is given by Eq.~(\ref{fulllikelihood}), while
$\mathcal{P}(\invhl)$ and $\mathcal{P}(\bii)$ are the prior PDFs for
\invhl\ and for the background indices, respectively.  The one-dimensional
posterior PDF
$\mathcal{P}(\invhl|\boldsymbol{\mathcal{D}},\boldsymbol{\theta})$ of the
parameter \invhl\ of interest is derived by marginalization over all nuisance
parameters {\bf BI}. The marginalization is performed by the BAT
toolkit~\cite{bat} via a Markov chain Monte Carlo numerical integration.  A
flat PDF between 0 and 0.1~\ctsper\ is considered as prior for all background
indices. As in Ref.~\cite{gerda:2013:prl}, a flat prior distribution is taken
for \invhl\ between 0 and $10^{-24}$\,/yr, i.e. all counting rates up to a
maximum are considered to be equiprobable.  The parameters
$\boldsymbol{\theta}$ in the likelihood $ \mathcal{L}$ are fixed during the
Bayesian analysis and the uncertainties are folded into the posterior PDF as
last step by an integral average:
\begin{equation}
   \langle \mathcal{P}(\invhl|\boldsymbol{\mathcal{D}}) \rangle = \int
   \mathcal{P}(\invhl|\boldsymbol{\mathcal{D}},\boldsymbol{\theta}) 
\prod_i g(\theta_i) d\theta_i
\end{equation}
with $g(\theta_i)$ being Gaussian distributions like for the frequentist
analysis.  The integration is performed numerically by a Monte Carlo approach.

The median sensitivity of the experiment in the case of no signal is
$\thalfzero>3.1\cdot 10^{25}$\,yr (90\% C.I.).  The posterior PDF
$\langle\mathcal{P}(\invhl|\boldsymbol{\mathcal{D}})\rangle$ for our data has
an exponential shape with the mode at $\invhl=0$. Its 90\,\% probability
quantile yields $\thalfzero > 3.5\cdot 10^{25}$\,yr.

As in any Bayesian analysis, results depend on the choice of the priors. For
our limit we assume all signal count rates to be a priori equiprobable.
Alternative reasonable choices are for instance: equiprobable Majorana
neutrino masses, which yields a prior proportional to $1/\sqrt{\invhl}$; or
scale invariance in the counting rate, namely a flat prior in
$\log(\invhl)$. The limits derived with these assumptions are significantly
stronger (50\,\% or more), since for both alternatives the prior PDFs increase
the probability of low \invhl\ values.

The systematic uncertainties weaken the limit on \invhl\ by less than 1\% both
in the frequentist and Bayesian analysis.  In general, the impact of
systematic uncertainties on limits is marginal in the low-statistics regime
that characterizes our experiment (see also Ref.~\cite{cousins}).

The limit derived from the \gerda\ data is slightly stronger than the median
sensitivity.  This effect is more significant in the frequentist analysis as
one would expect, see e.g. Ref.~\cite{biller} for a detailed discussion.  The
probability of obtaining a frequentist (Bayesian) limit stronger than the
actual one is 33\,\% (35\,\%).
%%%%%%%%%%%%%%%%%%%%%%%%%%%%%%%%%%%%%%%%%%%%%%%%%%%%%%%%%%%%%%%%%%%%%%%%%%%%%%%
\bibliography{gerda_paper}
\providecommand{\noopsort}[1]{}\providecommand{\singleletter}[1]{#1}%
%

%\clearpage

%%%%%%%%%%%%%%%%%%%%%%%%%%%%%%%%%%%%%%%%%%%%%%%%%%%%%%%%%%%%%%%%%%%%%%%%%%%%%%%
%\setcounter{figure}{0}
\setcounter{exfig}{1}   
\renewcommand{\figurename}{Extended Data Fig.}
\renewcommand\thefigure{\arabic{exfig}}    

\begin{figure*}
\includegraphics[width=0.9\textwidth]{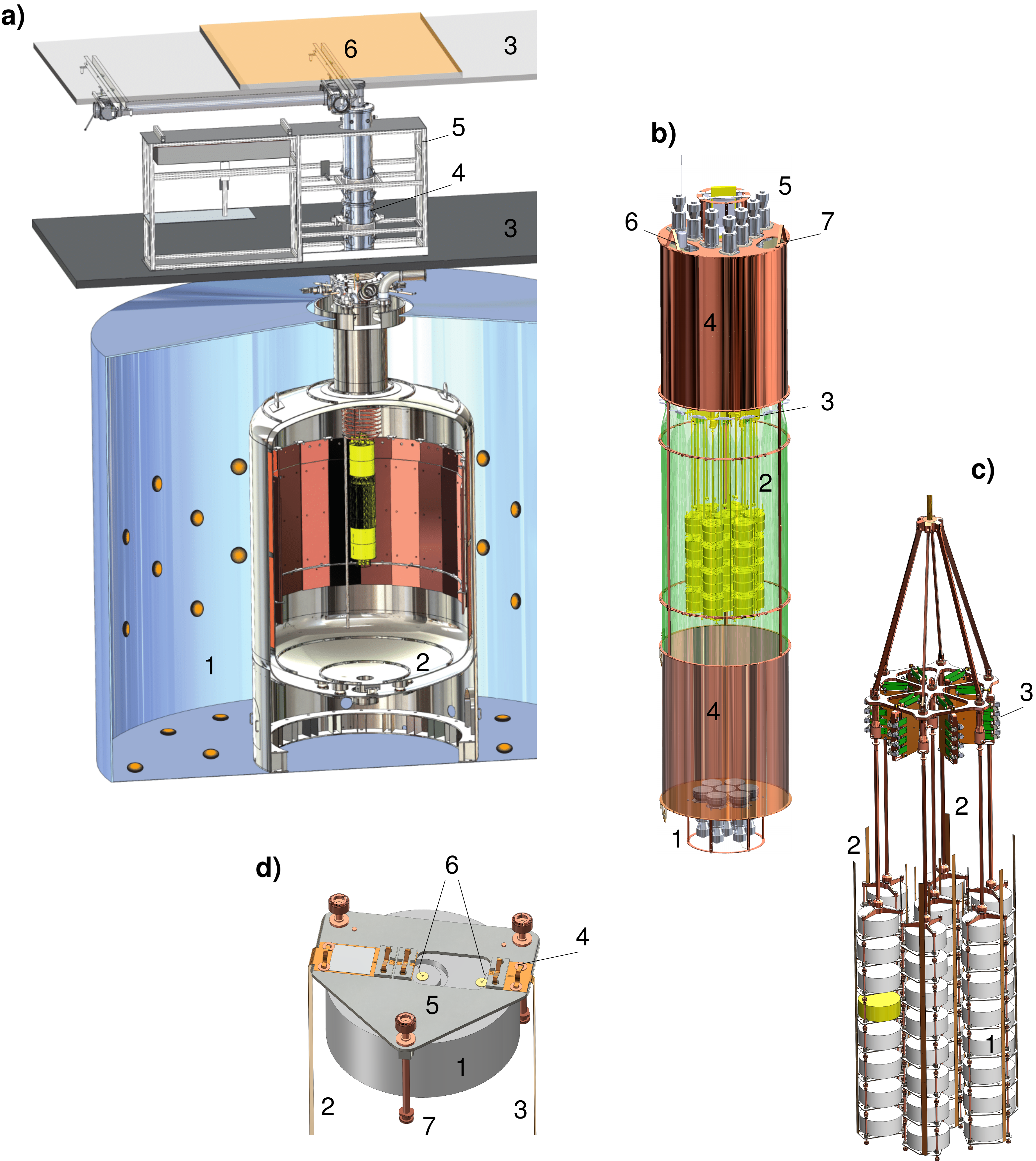}
\caption{\label{extfig:setup}
    \gerda~Phase~II experimental setup. \hfill \break {\it a) overview:} (1)
    water tank with muon veto system PMTs (590\,m$^3$, diameter\,10\,m), (2)
    LAr cryostat (64\,m$^3$, diameter\,4\,m,), (3) floor \& roof of clean
    room, (4) lock, (5) glove box (6) plastic muon veto system; \hfill \break
    {\it b) LAr veto system:} (1/5) bottom/top plate (diameter\,49\,cm) with
    7/9 3"~PMTs (R11065-10/20 MOD), low radioactivity of U and Th
    ($<$2\,mBq/PMT), (2) fiber curtain (h\,90\,cm) coated with wavelength
    shifter, (3) optical couplers and SiPMs, (4) thin-walled (0.1\,mm) Cu
    cylinders (h\,60\,cm) covered with a Tyvek reflector on the inside (6)
    calibration source entering slot in top plate (7) slot for second of three
    calibration sources; \hfill \break {\it c) detector array:} (1) Ge
    detectors arranged in 7 strings, (2) flexible bias and readout cables, (3)
    amplifiers; \hfill \break {\it d) detector module, view from bottom:} (1)
    BEGe diode (2/3) signal/high voltage cables attached by (4) bronze clamps
    to (5) silicon support plate, (6) bond wire connections from diode to
    signal and high voltage cable, (7) Cu support rods.
}
\end{figure*}
%\clearpage
\addtocounter{exfig}{1}   

\begin{figure*}
\includegraphics[width=10cm]{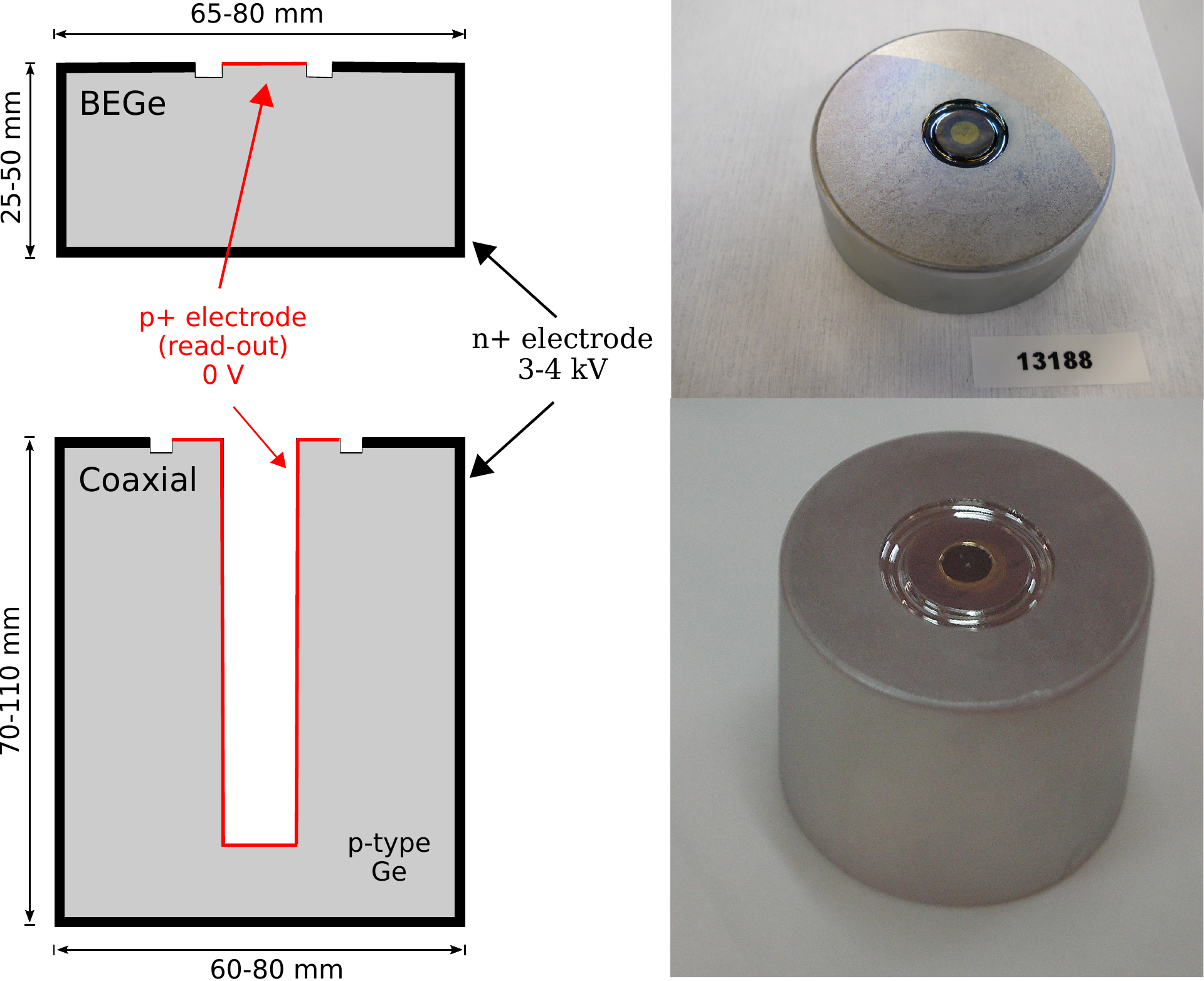}
\caption{\label{extfig:detectors}
    Cross section through the germanium detector types (left) and the
    corresponding photos (right).  The p$+$ electrode is made by a
    $\sim$0.3\,\,$\upmu$m thin boron implantation. The n$+$ electrode is a
    1-2~mm thick lithium diffusion layer and biased with up to +4500~V. The
    electric field drops to zero in the n$+$ layer and hence energy
    depositions in this fraction of the volume do not create a readout signal.
    The p$+$ electrode is connected to a charge sensitive amplifier.
}
\end{figure*}
%\clearpage
\addtocounter{exfig}{1}   

\begin{figure*}
\includegraphics[width=10cm]{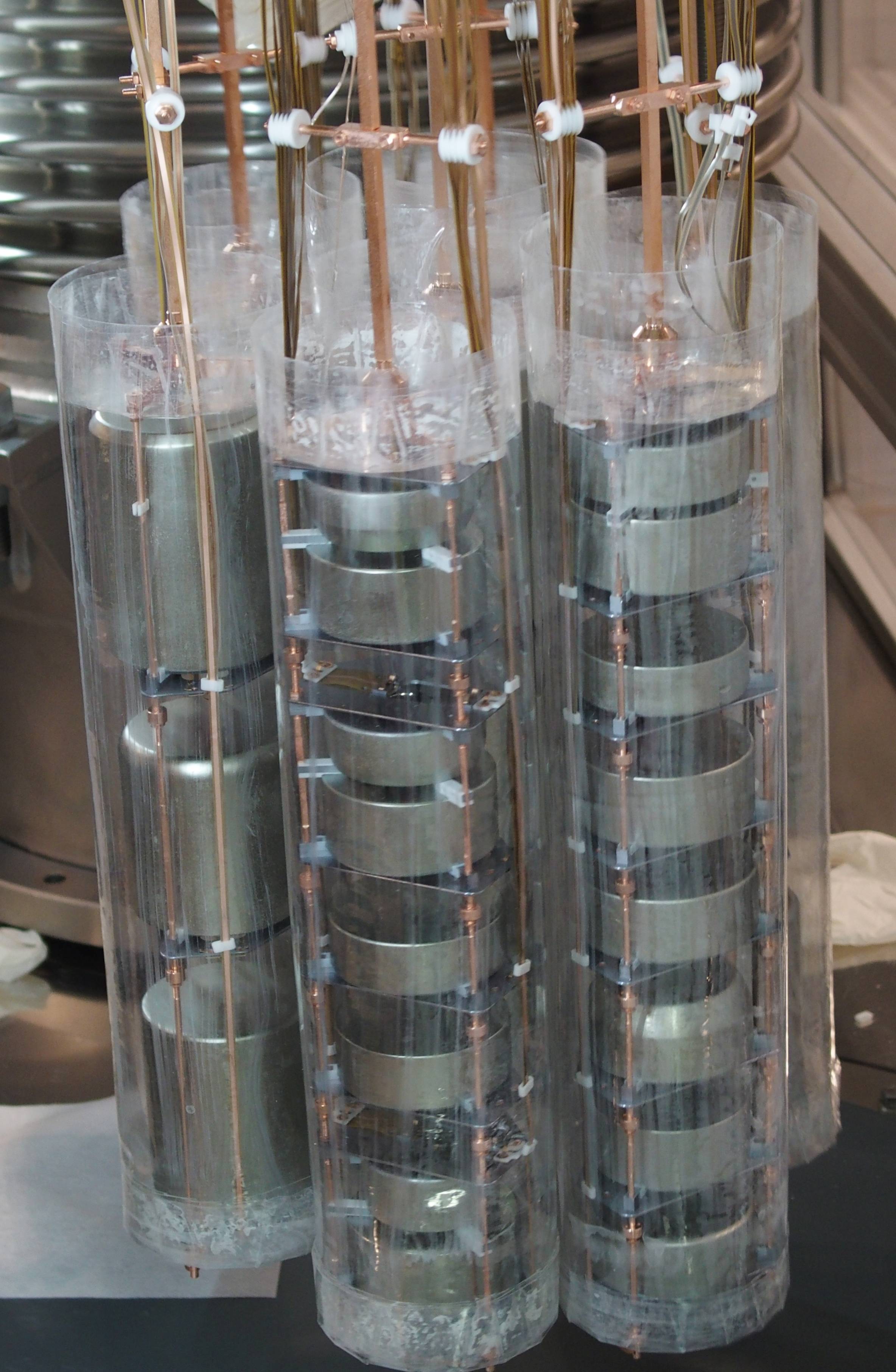}
\caption{\label{extfig:string}
          Photo of the assembled detector array with a string of coaxial
          detectors (left) and BEGe detectors (middle and right).  All are
          inside a shroud of nylon cylinders covered with a wavelength
          shifter.
}
\end{figure*}
%\clearpage
\addtocounter{exfig}{1}   

\begin{figure*}
\begin{minipage}{\columnwidth}
\includegraphics[width=\columnwidth]{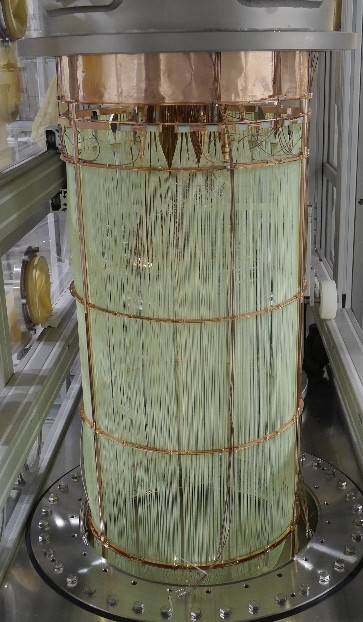}
\end{minipage}
\begin{minipage}{\columnwidth}
\includegraphics[width=\columnwidth]{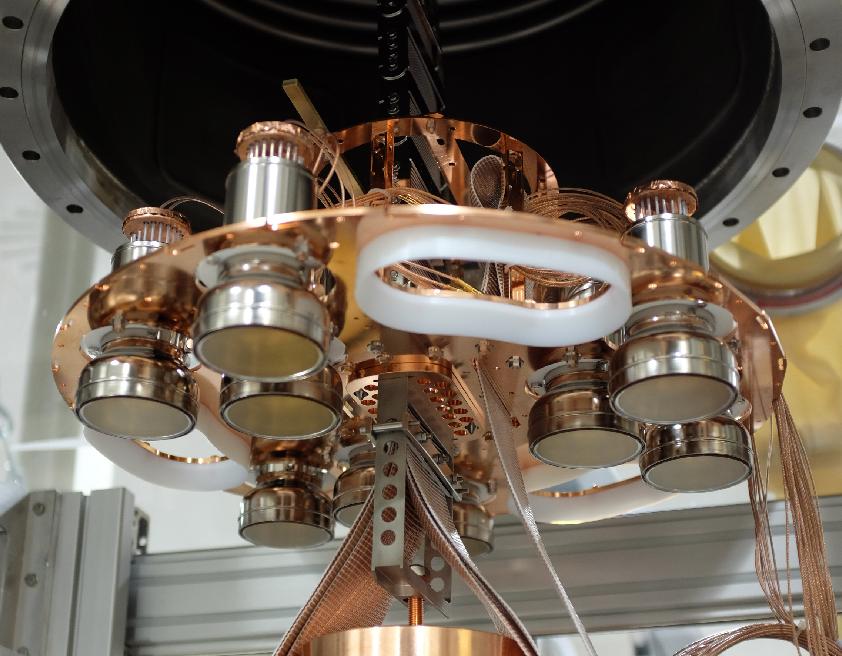}\\[5mm]
\includegraphics[width=\columnwidth]{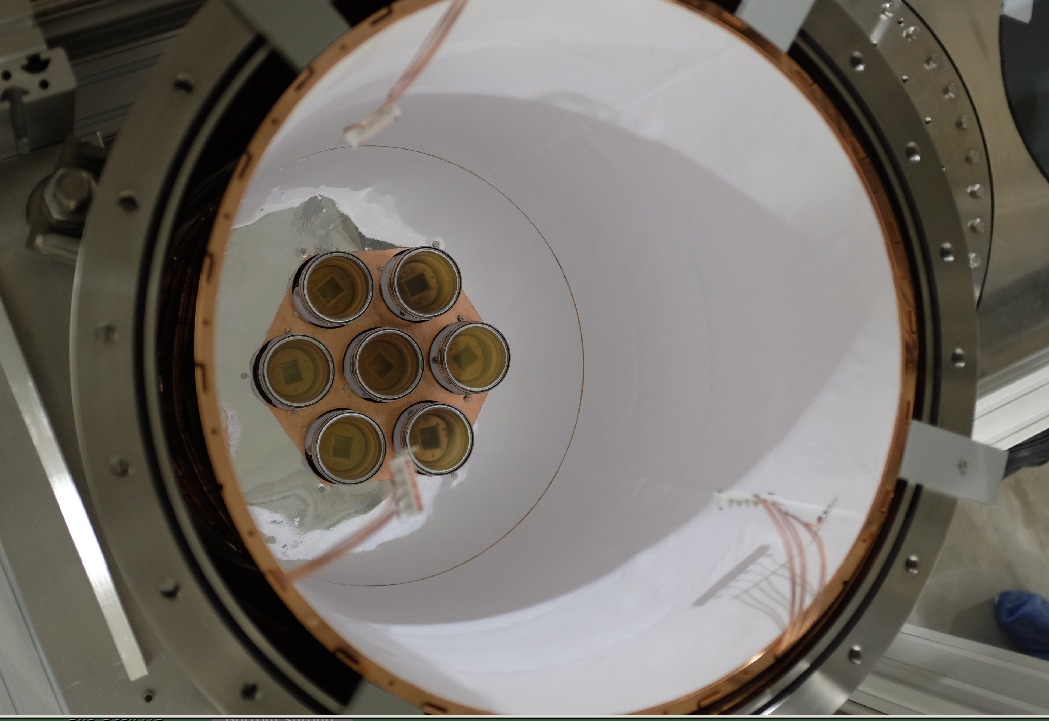}
\end{minipage}
\caption{\label{extfig:larcaps}
          Photos of the liquid argon veto system. Left: fiber curtain
          with SiPM readout at the top. Right: top and bottom arrangement of
          PMTs.
}
\end{figure*}
\clearpage
\addtocounter{exfig}{1}   

\begin{figure*}
\includegraphics[width=\textwidth]{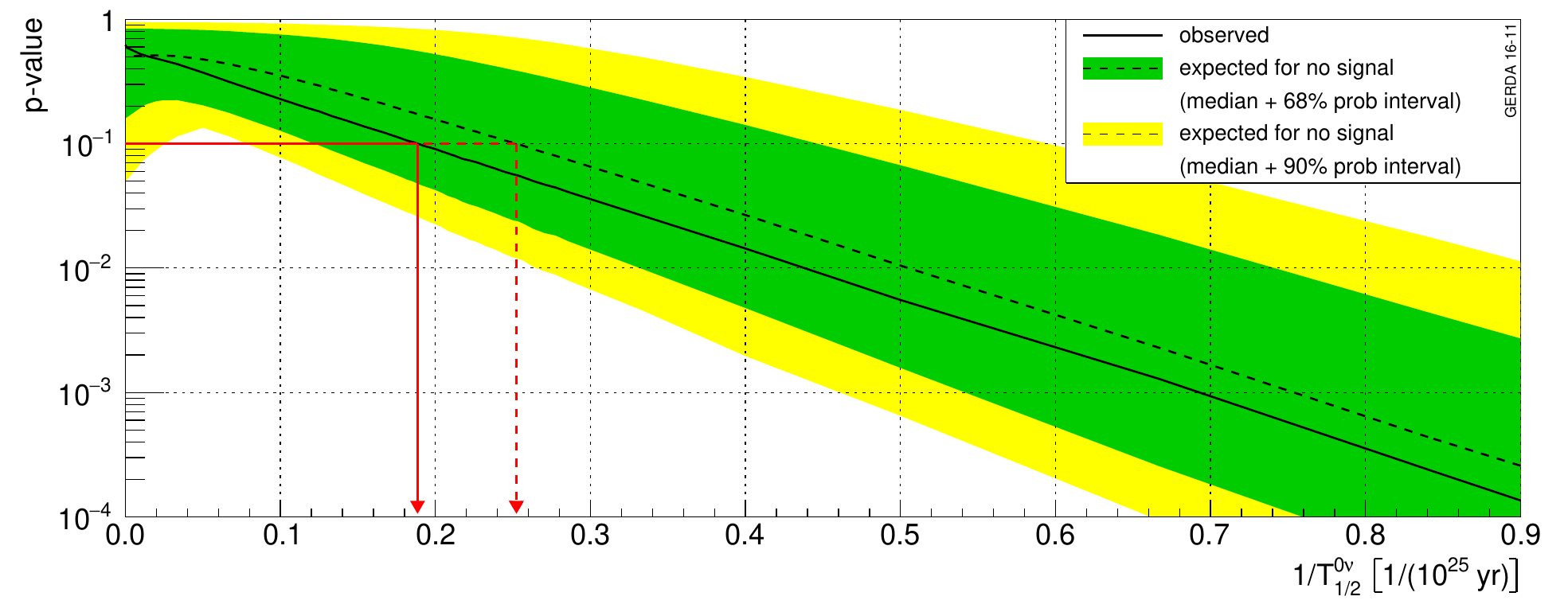}
\caption{\label{extfig:analysis}
          p-value for the hypothesis test as a function of the inverse
          half-life $1/T_{1/2}^{0\nu}$ for the data (full black line) and the
          median sensitivity (dashed black line). The 68 (90)\,\% interval is
          given by the green (yellow) range. The red arrows indicate the
          results at 90\,\% confidence level: the limit for $T_{1/2}^{0\nu}
          ({\rm ^{76}Ge})>5.3\cdot10^{25}$~yr (full red line), the median
          sensitivity for $T_{1/2}^{0\nu} ({\rm ^{76}Ge})>4.0\cdot10^{25}$~yr
          (dashed red line).
}
\end{figure*}

\end{document}